# Condensation of vortices in the X-Y model in 3d: a disorder parameter

G. Di Cecio[a] A. Di Giacomo[b] * G. Paffuti[b] M. Trigiante[c]

[a]Now at: Departement of Physics, Louisiana State University, Baton Rouge.

[b]Dipartimento di Fisica Università di Pisa and INFN Sezione di Pisa

[c]SISSA, Trieste

A disorder parameter is constructed which signals the condensation of vortices. The construction is tested by numerical simulations on lattice.

## 1. Introduction

The $XY$ model in $3d$ describes the critical behaviour of superfluid $He_4$[1]. It also provides a simple example of phase transition in which the condensation of solitons (specifically vortices), plays an essential role[2-5]. The phase transition is second order and the basic critical indices are known with good accuracy[6,9]. Viewed as the euclidean version of a $(2+1)$ dimensional quantum field theory, with the temperature $T$ as coupling constant, the system has a $U(1)$ symmetry describing the conservation of the number of vortices. Phenomenological analyses indicate that for $T > T_c$ this $U(1)$ is spontaneously broken, by condensation of vortices in the $2d$ ground state[9]. We will produce microscopic evidence for this phenomenon. We will construct the creation operator of a vortex, $\mu$, and use its v.e.v. $\langle\mu\rangle$ as a disorder parameter to detect condensation of vortices. A similar construction has been used to demonstrate the condensation of monopoles in compact $U(1)$[10] and in $SU(2)$ gauge theory as a mechanism for confinement of colour[11].

The action of the model is $(\beta = 1/T)$

$$S = \beta \sum_i \sum_{\mu=0}^{2} (1 - \cos(\Delta_\mu \theta(i))) \qquad (1)$$

The field variable is the angle $\theta$ at the site $i$.

At large $\beta$ the system describes a free massless particle

$$S \underset{\beta\to\infty}{\simeq} \frac{\beta}{2} (\Delta_\mu \theta)^2 \qquad (2)$$

For $\beta < \beta_c = .45419$ higher orders become important, and the density of vortices increases dramatically [3-5]. A vector field $A_\mu = \partial_\mu \theta$ can be defined and a current

$$j_\mu = \varepsilon_{\mu\nu\rho} \partial^\nu A^\rho \qquad (3)$$

For non singular configurations $j_\mu = 0$. By construction this current is conserved:$\partial_\mu j^\mu = 0$. The corresponding constant of motion is

$$\begin{aligned} V &= \int \mathrm{d}^2 x\, j^0(\vec{x}, x^0) = \int \mathrm{d}^2 x\, \varepsilon_{0ij} \partial^i A^j = \\ &= \oint_C \vec{A} \cdot \mathrm{d}\vec{x} = \oint_C \vec{\nabla}\theta \cdot \mathrm{d}\vec{x} \end{aligned} \qquad (4)$$

Single valuedness of the action implies that the last integral is an integer multiple of $2\pi$ or

$$V = n\, 2\pi \qquad (5)$$

$n$ is the number of vortices.

There exist configurations with non trivial $n$, e.g.

$$\bar{\theta}(x-y) = \arctan \frac{(x-y)_2}{(x-y)_1} \qquad (6)$$

which is singular at $\vec{x} = \vec{y}$ and has $n = 1$.

The creation operator of a vortex is the translation of $\theta$ by $\bar{\theta}$ or, being $\sin(\Delta_0 \theta)$ the conjugate momentum[10]

$$\mu(x) = \exp\left[ \mathrm{i} \int \mathrm{d}^2 y\, \bar{\theta}(\vec{x} - \vec{y}) \sin(\Delta_0 \theta(\vec{y})) \right] \qquad (7)$$

*Presented by A. Di Giacomo. Partially supported by MURST and by EC Contract CHEX-CT92-0051

A lattice (euclidean) version of $\mu$ is[10,12]

$$\mu(\vec{n},n_0) = \exp[-\beta \sum_{\vec{n}'} \{\cos\left(\Delta_0\theta(\vec{n}',n_0)\right.$$
$$\left.-\bar{\theta}(\vec{n}-\vec{n}')\right) - \cos(\Delta_0\theta(\vec{n}',n_0))\}] \qquad (8)$$

where $\vec{n}'$ runs on the slice $n_0$ = const., on all points of the lattice except the location $\vec{n}$ of the vortex.

We will compute $\langle\mu\rangle$ and show that it vanishes for $\beta > \beta_c$ (in the limit $V \to \infty$), and is $\neq 0$ for $\beta < \beta_c$: $\langle\mu\rangle$ is thus a disorder parameter and monitors the condensation of vortices.

## 2. Results.

**1)**

For large $\beta$'s $\langle\mu\rangle$ can be computed in perturbation theory, with the result

$$\langle\mu\rangle \simeq \exp\left[-\beta\left(c_1 V^{1/3} + c_2 + \mathcal{O}(1/\beta)\right)\right] \qquad (9)$$

$$c_1 = -11.332 \quad c_2 = 72.669$$

For $\beta > \beta_c$ and $V \to \infty$ $\langle\mu\rangle \to 0$.

**2)**

$\langle\mu\rangle$ can be computed from the correlation vortex - antivortex. At large distances, by use of cluster property and $C$ invariance,

$$\langle\mu(\vec{x},0)\bar{\mu}(\vec{y},t)\rangle \underset{t\to\infty}{\simeq} \langle\mu\rangle^2 \qquad (10)$$

Instead of measuring $\langle\mu\rangle$ directly, it proves convenient to measure $\rho$, defined as[10,11]

$$\rho = \frac{\mathrm{d}}{\mathrm{d}\beta}\ln\langle\mu\rangle \qquad (11)$$

In terms of $\rho$ $\mu = \exp\left(\int_0^\beta \rho(x)\mathrm{d}x\right)$. $\rho$ has a sharp negative peak around $\beta_c$, which signals a drop of $\langle\mu\rangle$ towards zero. (fig. 1).

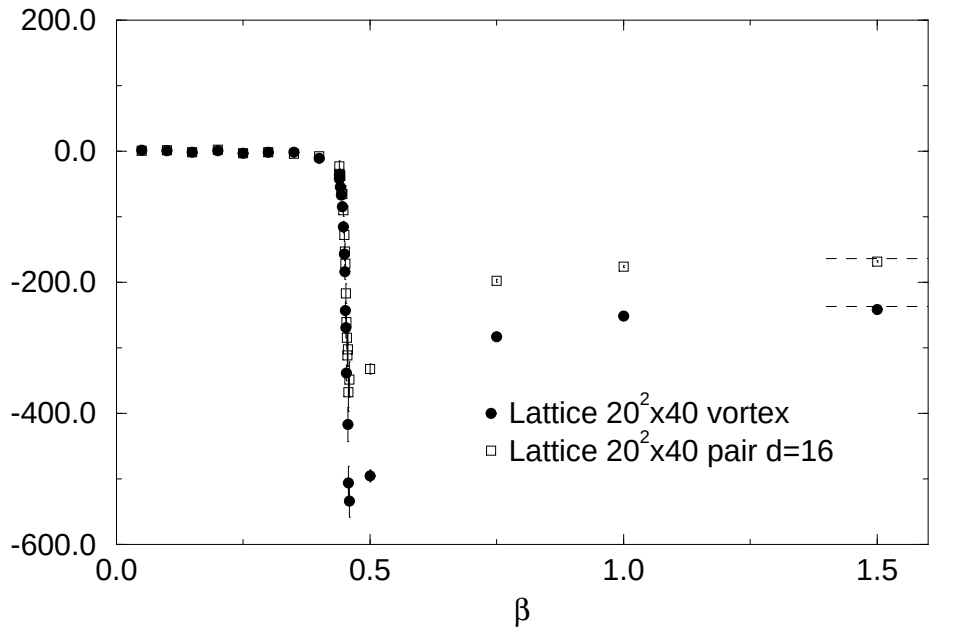

Figure 1. $\rho$ as a function of $\beta$. The dashed lines are the perturbative estimates at high $\beta$, Eq.(9).

For $\beta < \beta_c$ $\rho$ has a finite limit as $V \to \infty$ (fig. 2) implying that in this range $\langle\mu\rangle \neq 0$.

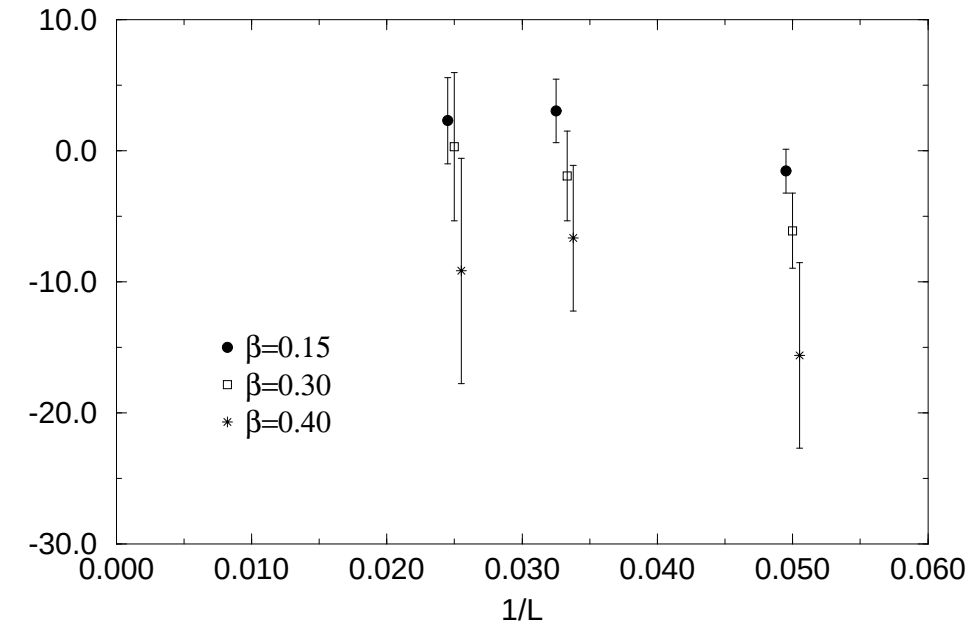

Figure 2. $\rho$ as a function of $1/L$ in the condensed phase.

Around $\beta_c$ a finite size scaling analysis can be performed as follows. If $\xi$ is the correlation length,

$$\xi \underset{\beta\to\beta_c^-}{\simeq} (\beta_c - \beta)^{-\nu}$$

$$\langle\mu\rangle = \langle\mu\rangle\left(\frac{\xi}{L},\frac{a}{L}\right) \underset{\beta\to\beta_c^-}{\simeq} \langle\mu\rangle\left(\frac{\xi}{L},0\right) \qquad (12)$$

or

$$\langle\mu\rangle = \langle\mu\rangle\left(L^{1/\nu}(\beta_c-\beta)\right) \qquad (13)$$

and

$$\rho = L^{1/\nu} f\left(L^{1/\nu}(\beta_c-\beta)\right) \qquad (14)$$

The quality of scaling law, Eq.(14), is shown in fig. 3. A best fit to the data[12] for $L = 20, 30, 40$ gives, with $\chi^2/dof = 1.07$

$$\beta_c = .4538(3) \quad \nu = .669(65)$$

to be compared to

$$\beta_c = .45419(2) \quad \nu = .670(7)$$

of [8].

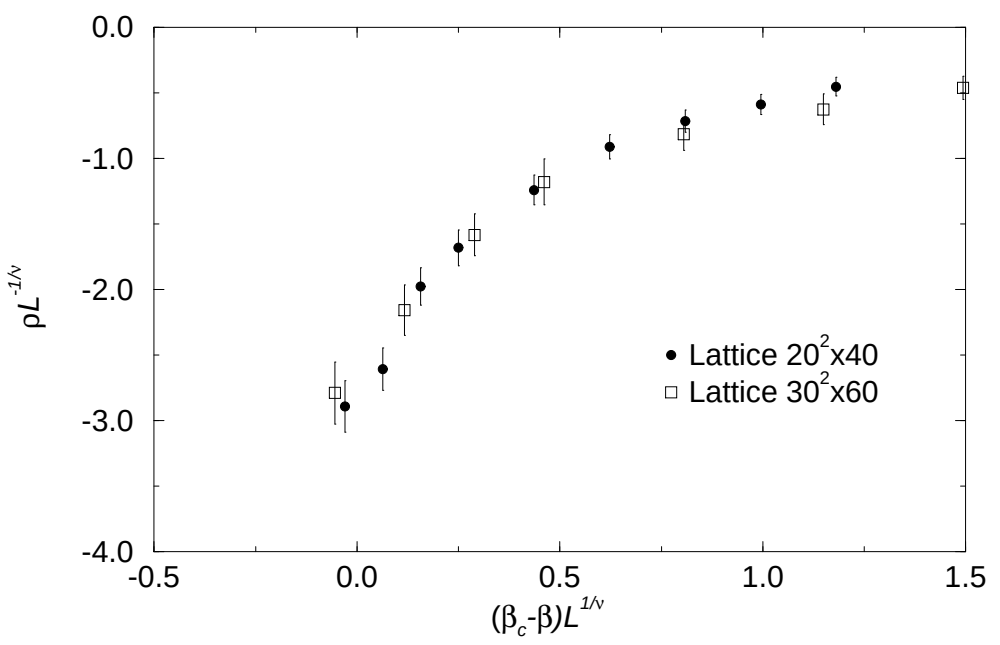

Figure 3. Quality of the finite size scaling analysis. The figure corresponds to the optimal values of $\beta_c$ and $\nu$.

The critical index for $\langle\mu\rangle$, ($\langle\mu\rangle \sim_{\beta\to\beta_c^-} (\beta_c-\beta)^\delta$) is $\delta = .740(29)$.

**3)**

The form (8) for $\mu(\vec{n})$ gives for $\langle\mu(\vec{n},n_0)\bar{\mu}(\vec{m},m_0)\rangle$

$$\langle\mu\bar{\mu}\rangle = \frac{1}{Z}\int\prod\frac{\mathrm{d}\theta_i}{2\pi}\exp(-S-S') \qquad (15)$$

with

$$Z = \int\prod\frac{\mathrm{d}\theta_i}{2\pi}\exp(-S)$$

$S$ is defined by Eq.(1), $S' = \ln\mu + \ln\bar{\mu}$ by Eq.(8). $S + S'$ is nothing but the replacement of the term $\cos(\Delta_0\theta(\vec{n}',n_0))$ in the action by $\cos(\Delta_0\theta(\vec{n}',n_0) + \bar{\theta}(\vec{n}-\vec{n}'))$ on the time slice $n_0$ where the vortex is created and a similar replacement at time $m_0$ where the vortex is destroyed. Since all the $\theta_i$ in Eq.(15) appear as arguments of periodic functions, any change of variables $\theta_i \to \theta_i + f_i$ ($A_\mu \to A_\mu + \Delta_\mu f$) leaves $\langle\mu\rangle$ invariant.

A change of variables

$$\theta(\vec{n}',n_0+1) \to \theta(\vec{n}',n_0+1) + \bar{\theta}(\vec{n}-\vec{n}')$$

sends

$\cos(\theta(\Delta_0\theta(\vec{n},n_0) - \bar{\theta}(\vec{n}-\vec{n}')) \to \cos(\theta(\Delta_0\theta(\vec{n},n_0))$ and $\cos(\theta(\Delta_0\theta(\vec{n},n_0+1)) \to \cos(\theta(\Delta_0\theta(\vec{n},n_0+1) - \bar{\theta}(\vec{n}-\vec{n}'))$ On the slice $n_0+1$ the boundary conditions change and the number of vortices is changed by the number of vortices $\bar{n}$ carried by $\bar{\theta}$. A similar change of variables can again be performed on the slice $n_0+2$ where again the number of vortices gets changed by $\bar{n}$, and so on, until the time $m_0$ is reached where the vortex is destroyed. From that time on the the change of variables is by $\bar{\theta}(\vec{n}-\vec{n}') - \bar{\theta}(\vec{m}-\vec{m}')$ which carries number of vortices zero. Hence the operator $\mu$ properly changes the boundary conditions when monopoles are created or destroyed.

In conclusion we have defined a good disorder parameter for condensation of vortices. It describes physics correctly. It also provides a good test of the procedure used for detecting condensation of monopoles in gauge theories[10-11].

## REFERENCES

1. R.L. Onsager: *Nuovo Cimento Suppl.* **6**, 249, (1949); R.P. Feynman: in *Progress in Low Temperature Physics,* C.J. Gorter ed., North Holland, Amsterdam (1955), Vol. 1, p.17.
2. T. Banks, R. Meyerson, J.B. Kogut: *Nucl. Phys* **B129**, 493, (1977).
3. G. Kohring, R.E. Shrock, P. Wills: *Phys. Rev. Lett. 57*, 1358, (1986).
4. S.R. Shenoy: *Phys. Rev.* **B40**, 5056, (1989).
5. M. Ferer, M.A. Mosre, M. Wortiz: *Phys. Rev.* **B8**, 5205, (1973).
6. W. Janke, H. Kleinert: *Nucl. Phys.* **B270**, 399, (1986).
7. W. Janke: *Phys. Lett.* **A148**, 306, (1990).
8. A.P. Gottlob, M. Hasenbusch: *CERN TH* 6885-93.
9. H. Kleinert: *Phys. Lett.* **93 A**, 86, (1982).
10. L. Del Debbio, A. Di Giacomo, G.Paffuti: *Phys. Lett.* **B349**, 513, (1995).
11. L. Del Debbio, A. Di Giacomo, G.Paffuti, P. Pieri: *Phys. Lett.* **B355**,255, (1995).
12. G. Di Cecio, A. Di Giacomo, G. Paffuti, M. Trigiante: Pisa preprint IFUP-TH 13/96 and cond-mat 9603139